\documentclass[useAMS,usenatbib,usegraphicx]{mn2e}

\title[SZ power spectrum with decaying cold dark matter]
{Sunyaev-Zel'dovich power spectrum with decaying cold dark matter}
\author[K. Takahashi, M. Oguri and K. Ichiki]
{Keitaro Takahashi$^{1}$\thanks{E-mail:ktaro@utap.phys.s.u-tokyo.ac.jp},
Masamune Oguri$^{1}$\thanks{E-mail:oguri@utap.phys.s.u-tokyo.ac.jp}
and Kiyotomo Ichiki$^{2,3}$\thanks{E-mail:kiyotomo.ichiki@nao.ac.jp}\\
$^{1}$Department of Physics, School of Science, University of Tokyo, 
Hongo 7-3-1, Bunkyo-ku, Tokyo 113-0033, Japan\\
$^{2}$Department of Astronomy, School of Science, University of Tokyo, 
Hongo 7-3-1, Bunkyo-ku, Tokyo 113-0033, Japan\\
$^{3}$Division of Theoretical Astrophysics, National
Astronomical Observatory, 2-21-1 Osawa, Mitaka, Tokyo 181-8588, Japan}
\begin{document}


\pagerange{\pageref{firstpage}--\pageref{lastpage}} \pubyear{2003}

\maketitle

\label{firstpage}

\begin{abstract}
 Recent studies of structures of galaxies and clusters imply that dark 
 matter might be unstable and decay with lifetime $\Gamma^{-1}$ about the
 age of universe. We study the effects of the decay of cold dark matter on
 the Sunyaev-Zel'dovich (SZ) power spectrum. We analytically calculate 
 the SZ power spectrum taking finite lifetime of cold dark matter into
 account. We find the finite lifetime of dark matter decreases the power
 at large scale ($l < 4000$) and increases at small scale ($l > 4000$). 
 This is in marked contrast with the dependence of other cosmological
 parameters such as the amplitude of mass fluctuations $\sigma_{8}$ and
 the cosmological constant $\Omega_{\lambda}$ (under the assumption of a
 flat universe) which mainly change the normalization of the angular
 power spectrum. This difference allows one to determine the lifetime
 and other cosmological parameters rather separately. We also
 investigate sensitivity of a future SZ survey to the cosmological
 parameters including the life time, assuming a fiducial model
 $\Gamma^{-1} = 10 h^{-1} {\rm Gyr}$, $\sigma_{8} = 1.0$, and
 $\Omega_{\lambda} = 0.7$. We show that future SZ surveys such as ACT,
 AMIBA, and BOLOCAM can determine the lifetime within factor of two even
 if $\sigma_{8}$ and $\Omega_{\lambda}$ are marginalized. 
\end{abstract} 

\begin{keywords}
cosmological parameters --- cosmology: theory --- dark matter
--- cosmic microwave background --- galaxies: clusters: general
\end{keywords}

\section{INTRODUCTION}

The $\Lambda$-dominated Cold Dark Matter (${\rm \Lambda CDM}$) model,
in which universe is dominated by cold dark matter and dark energy, 
is remarkably successful in many ways. However, recent analyses of
structure on galactic and sub-galactic scales have suggested
discrepancies and stimulated numerous alternative proposals
\citep{ostriker03}. Decaying cold dark matter is one of such proposals;
\citet{cen01} investigated the effect of cold dark matter decay on the
density profile of a halo and found that it can solve problems of
overproduction of small dwarf galaxies as well as overconcentration of
galactic halos in the ${\rm \Lambda CDM}$ model if half of the dark
matter particles decay into relativistic particles by $z=0$. Also
decaying dark matter which is superheavy ($> 10^{12}$GeV) has been
proposed as the origin of ultra-high energy cosmic rays
\citep{kuzmin98,hagiwara01}. \citet{chou03} showed that such a dark
matter with lifetime about the age of the universe can solve both the
puzzle of ultra-high energy cosmic rays and the problem of
overproduction of small dwarf galaxies simultaneously. 

The effect of decaying cold dark matter would be seen also in more larger
scale: clusters of galaxies and the universe itself. \citet{ichiki03}
showed that decaying cold dark matter with lifetime about the age of
universe improves the fit of Type Ia supernova observations, the 
evolution of mass-to-light ratios and the fraction of X-ray emitting gas
of clusters, and cosmic microwave background (CMB) observations. From
these observations, they set a constraint on the lifetime $\tau$, $15 <
\tau < 80$Gyr. Furthermore, \citet{oguri03} developed a method to
compute the mass function of clusters in decaying cold dark matter model
on the basis of the Press-Schechter formalism \citep{press74}. They
found that a finite lifetime of cold dark matter significantly changes
the evolution of the cluster abundance and that the observed evolution
of the cluster abundance can be accounted well if the life time is about
the age of the universe. 

It is interesting that all these studies suggest the lifetime of cold
dark matter to be the age of the universe. Thus such decaying cold dark
matter should be investigated more rigorously. In this paper, we study
the effect of finite lifetime of cold dark matter on power spectrum of
Sunyaev-Zel'dovich (SZ) effect and investigate the possibility to probe
the lifetime of dark matter by future CMB experiment. Finite lifetime
changes the expansion law of the universe and the evolution of density
fluctuation, both of which affect the SZ power spectrum.

Theoretical candidates for decaying cold dark matter have been proposed
by many authors and their predictions for lifetime of such particles
cover a large range of values $10^{-2} < \tau < 10^{11}$ Gyr
\citep{Chung98,Benakli98,Kim:2001sh}. Decaying dark matter with lifetime
around the age of the universe has significant observational signals if
it decays into observable particles. The most stringent constraint may
come from the diffuse gamma ray background observations
\citep{Dolgov:1981hv}. However, realistic calculation which takes all
energy dissipation processes into account showed that even the particles
with lifetime as short as a few times of the age of the universe still
are not ruled out by recent observations \citep{Ziaeepour:2000rc}.
Moreover, as long as we have not identified what dark matter is,
the decay channel cannot escape some uncertainties. On the contrary,
cosmological constraints such as those from CMB and SZ effect do not
depend on the details of the decay products. Here we assume only that
dark matter particles decay into relativistic particles. 

Recently, dark matter with finite lifetime has been renewed as
disappearing dark matter in the context of the brane world scenario 
\citep{randall99a,randall99b}. In the brane world scenario a bulk scalar
field can be trapped on the brane, which is identified with our universe,
and become cold dark matter. But since the scalar field is expected to be
metastable, it decays into continuum states in the higher dimension with
some decay width $\Gamma$, which is determined by the mass of the scalar 
field and the energy scale of the extradimension \citep{dubovsky00}. 
An observer on the brane sees this as if the scalar field disappears into 
the extradimension leaving some energy on the brane. This energy behaves 
in the same way as energy of relativistic particles but does not
correspond to real particles. Thus this is called ``dark radiation''. 
Therefore, this model cannot be constrained from decay products such as
diffuse gamma ray background; one needs cosmological constraints to test
this model. Since they contribute to the expansion of the universe in
the same way, hereafter we call both of them decaying cold dark matter. 

Let's turn to the ordinary matter which produce SZ effect. Our universe
is full of ionized gas. Hot electron scatters off CMB photons and
generates secondary CMB temperature fluctuations. This effect is SZ
effect \citep{sunyaev72} and has been rigorously studied as a powerful
tool for probing both cluster physics and cosmology. For recent review,
see \citet{review}.  

SZ effect reflects the characteristics of the cluster gas: the intensity
of the thermal SZ effect is proportional to the line-of-sight integral of
electron density times temperature. Since this dependence is different from
that of the X-ray thermal bremsstrahlung emission, combining these
observations allows us to study cluster physics effectively
\citep{zaroubi98,hughes98,yoshikawa99,gomez03}. Especially recent
high-resolution SZ image paved the way for a detailed study of complex
structures in the intracluster medium \citep{kitayama03}.

On the other hand, a survey of SZ effect reflects the number density
of clusters and its evolution. Since the number density
and its evolution are quite sensitive to some of cosmological
parameters, such as $\Omega_{m0}$, current density parameter of matter   
and $\sigma_{8}$, amplitude of mass fluctuations on a scale of $8
h^{-1}$Mpc, SZ survey can be used to constrain these parameters 
\citep{makino93,komatsu99,dasilva00,dasilva01a,seljak01,zhang01,AMIBA,komatsu02,battye03}.
Also statistical strategy to determine the cluster luminosity function
was developed in \citet{lee02} and \citet{lee03}.

Several CMB experiments such as CBI \citep{CBI1,CBI2}, BIMA \citep{BIMA}
and ACBAR \citep{ACBAR} detected the excess power over the primary
fluctuation and these may be the first SZ effect detections in a blank
sky survey. Furthermore, there are many upcoming CMB experiments such as
ACT \citep{komatsu02}, AMIBA \citep{AMIBA}, BOLOCAM \citep{BOLOCAM} and
Planck \citep{Planck}, which are expected to be able to measure the SZ
power spectrum with about $1\%$ accuracy. Thus precision era of SZ
effect is not far away. An accurate theoretical understanding required
for the future experiment has been proceeded 
by many authors \citep{oh03,diego03,zhang03a,zhang03b}.

The structure of this paper is as follows.
In section 2 and 3, we briefly review the phenomenology of decaying
cold dark matter and the calculation method of SZ power spectrum,
respectively. We give power spectrum taking
finite lifetime of dark matter into account and investigate the sensitivity
of future CMB experiment on lifetime and other cosmological parameters
in section 4. Finally, we summarize our results in section 5.
Throughout the paper, we assume a flat universe which is favored by
recent observations \citep[e.g.,][]{spergel03}.

\section{PHENOMENOLOGY OF DECAYING DARK MATTER}

Finite lifetime affects not only the expansion law of the universe
but also the evolution of density fluctuation. Here we review the method
to calculate the expansion law and the mass function. For details of
derivation and computation, see \citet{oguri03}.

\subsection{Cosmology}

We assume that dark matter particles decay into relativistic particles
and that the radiation component consists only of the decay product of
the dark matter. Then rate equations of matter and radiation components are,
\begin{eqnarray}
\dot{\rho}_{m} + 3 H \rho_{m} & = & - \Gamma \rho_{m}, \\
\dot{\rho}_{r} + 4 H \rho_{r} & = & \Gamma \rho_{m},
\end{eqnarray}
where $\rho_{m}$ and $\rho_{r}$ are energy density of matter and radiation,
respectively, dot denotes time derivative, and $\Gamma$ is the decay width 
of the dark matter. From these equations, we obtain the evolution of 
the matter and radiation energy density as,
\begin{eqnarray}
\rho_{m} & = & \rho_{m0} a^{-3} e^{-\Gamma t}, \\
\rho_{r} & = & \Gamma \rho_{m0} a^{-4} \int_{0}^{t} a e^{-\Gamma t} dt,
\end{eqnarray}
from which Friedmann equation is obtained as,
\begin{eqnarray}
\frac{H^{2}(a)}{H_{0}^{2}} 
& = & \Omega_{m0} a^{-3} e^{- \Gamma(t - t_{0})} \nonumber \\
& &
+ \Gamma \Omega_{m0} a^{-4} \int_{0}^{t} a e^{- \Gamma(t - t_{0})} dt
+ \Omega_{\lambda}.
\end{eqnarray}
Here $\Omega_{m0}$ and $\Omega_{\lambda}$ are the current density parameters
of the dark matter and dark energy, respectively. Combining this equation 
and the definition of the cosmological time,
\begin{equation}
t = \int_{0}^{a} \frac{da}{a H(a)},
\end{equation}
we obtain the following equation:
\begin{equation}
t'' = - \frac{1}{2} 
\left(\frac{\Omega_{m0} e^{- \Gamma(t - t_{0})}}{a^{2}}
      + 4 a \Omega_{\lambda} \right)
t'{}^3 + \frac{t'}{a},
\label{eq:cosmology}
\end{equation}
where prime denotes derivative with respect to the scale factor $a$.
It should be noted that the Friedmann equation at $z=0$ reduces to,
\begin{equation}
1 = \Omega_{m0} 
  + \Gamma \Omega_{m0} \int_{0}^{t_{0}} a e^{- \Gamma(t - t_{0})} dt
  + \Omega_{\lambda}.
\end{equation} 
Thus, the current density parameter $\Omega_{m0}$ is uniquely determined
by $\Gamma$ and $\Omega_{\lambda}$.

\subsection{Mass function}

In our previous paper \citep{oguri03}, we calculated mass function on
the basis of the Press-Schechter theory \citep{press74}. We first
consider motion of a spherical overdensity with radius $R$ and initial
mass $M$. The equation of motion of the spherical shell is given by,
\begin{equation}
\frac{\ddot{R}}{R} = 
- \frac{GM}{R^3} e^{-\Gamma t} -
 H_{0}^{2} \Gamma \Omega_{m0} a^{-4} \int_{0}^{t} a e^{-\Gamma(t-t_0)} dt
+ H_{0}^{2} \Omega_{\lambda}.
\label{eq:EOM}
\end{equation}
From equations (\ref{eq:cosmology}) and (\ref{eq:EOM}), we can calculate
the nonlinear overdensity $\Delta_{\rm c}$ and the extrapolation of the
linear fluctuation $\delta_{\rm c}$ at virialization, although we used
$\delta_{\rm c} = 1.58$ for simplicity. It can be interpreted that a
region has already been virialized at $z$ if the linearly extrapolated
density contrast  $\delta_{\rm linear}(M_{i},z)$, which is smoothed over
the region containing mass $M_{i}$, exceeds the critical value
$\delta_{\rm c}$. The evolution of the linear density contrast is
determined by, 
\begin{equation}
\dot{a}^2\delta'' 
+ \left( \ddot{a} + 2 \frac{\dot{a}^{2}}{a} \right)\delta'
- 4 \pi G \bar{\rho}_{m} \delta = 0.
\label{eq:linear}
\end{equation}
If the initial density field is random Gaussian, probability distribution
function of the density contrast is given by,
\begin{equation}
P \left[ \delta(M_{i}, z_{i}) \right] =
\frac{1}{\left( 2\pi \right)^{1/2} \sigma_{M_{i}}(z_{i})} 
\exp\left[
          - \frac{\delta^{2}(M_{i}, z_{i})}{2 \sigma^{2}_{M_{i}}(z_{i})}
    \right],
\end{equation}
where $\sigma_{M_{i}}(z_{i})$ is the mass variance. We assume the mass
variance for the cold dark matter fluctuation spectrum with the
primordial spectral index $n=1$, and adopt a fitting formula presented
by \citet{kitayama96} in which they used transfer function of
\citet{bardeen86}. Consequently, the probability that the region with
mass $M$ has already been virialized is obtained as,
\begin{equation}
f(M, t) = \frac{1}{2} {\rm erfc}
\left( \frac{\delta_{\rm c}(z)}{\sqrt{2} \sigma_{M_{i}}} \right),
\label{eq:erfc}
\end{equation}
where ${\rm erfc}(x)$ is the complementary error function,
$\sigma_{M_{i}} \equiv \sigma_{M_{i}}(z=0)$, and 
$\delta_{\rm c}(z) = \delta_{\rm c} D(z=0)/D(z)$. Here $D(z)$ is linear
growth rate, which can be calculated from equation (\ref{eq:linear}).
From this equation, we finally obtain the comoving number
density of halos of mass $M$ at redshift $z$,
\begin{eqnarray}
\frac{dn_{\rm PS}}{dM}(M, z) & = &
e^{2 \Gamma t} \sqrt{\frac{2}{\pi}} \frac{\rho_{0}}{M_{i}}
\frac{\delta_{\rm c}(z)}{\sigma_{M_{i}}^2}
\left| \frac{d\sigma_{M_{i}}}{dM_{i}} \right| \nonumber \\
& & \times \left.
\exp\left[ -\frac{\delta_{\rm c}^2(z)}{2\sigma_{M_{i}}^2} \right]
\right|_{M_{i}=M e^{\Gamma t}},
\label{eq:mass_function}
\end{eqnarray}
where $\rho_{0} = \rho_{\rm crit}(z=0) \Omega_{m0} e^{-\Gamma(t-t_{0})}$.
We use this mass function to compute the angular power spectrum of
SZ effect. Although the mass function of Press \& Schechter is claimed
to tend to underestimate the abundance of the massive halos
compared with $N$-body simulations \citep{jenkins01}, 
we assume that the dependence of the mass function on cosmological
parameters is well described by the Press-Schechter mass function.

\section{SUNYAEV-ZEL'DOVICH POWER SPECTRUM}

To compute the angular power spectrum of SZ effect, we basically follow 
\citet{komatsu02}. They derived a analytic prediction for the angular 
power spectrum using the universal gas-density and temperature profile 
which were developed in \citet{komatsu01}. Here we briefly review their
method. 

\subsection{Angular power spectrum}

Since for the angular scales of interest here, $l > 300$, the halo-halo
correlation term can be neglected, we consider only the one-halo Poisson
term. Then angular power spectrum is given by
\begin{eqnarray}
C_{l} & = & 
    g_{\nu}^{2} \int_{0}^{z_{\rm max}} dz \frac{dV(z,\Gamma)}{dz} \nonumber \\
& & \times \int_{M_{\rm min}}^{M_{\rm max}} dM
    \frac{dn_{\rm PS}(M,z,\Gamma)}{dM} |\tilde{y}_{l}(M,z,\Gamma)|^{2},
\end{eqnarray}
where $g_{\nu}$ is the spectral function of the SZ effect. $V(z,\Gamma)$ is
the comoving volume of the universe and this depends on $\Gamma$ because
expansion law depends on the lifetime of dark matter.  Here we take 
$z_{\rm max} = 10$, $M_{\rm min} = 10^{12} M_{\odot}$ and
$M_{\rm max} = 10^{16} M_{\odot}$, which were shown to be sufficient 
by \citet{komatsu02}. The 2D Fourier transform of the projected 
Compton $y$-parameter $\tilde{y}_{l}(M,z,\Gamma)$ is given by,
\begin{equation}
\tilde{y}_{l}(M,z,\Gamma) 
= \frac{4 \pi r_{\rm s}}{l_{s}^{2}}
  \int_{0}^{x_{\rm vir}} dx x^{2} y_{\rm 3D}(x) 
  \frac{\sin{l x/l_{\rm s}}}{l x/l_{\rm s}},
\label{eq:Compton_y}
\end{equation}
where $y_{\rm 3D}(x)$ is the 3D radial profile of the Compton $y$-parameter
and $x$ is a scaled, non-dimensional radius,
\begin{equation}
x \equiv r/r_{\rm s}.
\end{equation}
Here $r_{\rm s}$ is a scale radius which characterizes the 3D radial
profile. The corresponding angular wave number is
\begin{equation}
l_{s} \equiv d_{\rm A}/r_{\rm s},
\end{equation} 
where $d_{\rm A} = d_{\rm A}(z,\Gamma)$ is the proper angular-diameter 
distance. The scale radius $r_{\rm s}$ is parameterized by the
concentration parameter $c$:
\begin{equation}
c(M,z,\Gamma) \equiv \frac{r_{\rm vir}(M,z,\Gamma)}{r_{\rm s}(M,z)}
\sim \frac{10}{1+z} 
\left( \frac{M}{M_{*}(0)} \right)^{-0.2},
\end{equation}
where $r_{\rm vir}$ is the virial radius of halos and $M_{*}(0)$ is the
mass collapsing at redshift $z=0$, defined by $\sigma_M(z) = \delta_{\rm
c} \equiv 1.68$. The last expression follows from \citet{seljak00} and
we assume that this relation does not depend on $\Gamma$. The virial
radius $r_{\rm vir}$ can be calculated based on the spherical model,  
\begin{equation}
r_{\rm vir}(M,z,\Gamma) \equiv \left[
\frac{3 M}{4 \pi \Delta_{\rm c}(z,\Gamma) \rho_{0}(z)}
\right]^{1/3}.
\end{equation}
For the upper integration boundary of $x$ in equation
(\ref{eq:Compton_y}), we take a value $x_{\rm vir}$, which corresponds
to the virial radius $r_{\rm vir}$.  

The 3D radial profile of the Compton $y$-parameter $y_{\rm 3D}(x)$ is
written by a thermal gas-pressure profile $P_{\rm gas}(x)$, through,
\begin{eqnarray}
y_{\rm 3D}(x) & \equiv & \frac{\sigma_{\rm T}}{m_{e} c^{2}} P_{e}(x) 
\nonumber \\
& = & \frac{\sigma_{\rm T}}{m_{e} c^{2}} \frac{2 + 2X}{3 + 5X} P_{\rm gas}(x)
\end{eqnarray}
where $P_{e}(x)$ is an electron-pressure profile, $\sigma_{\rm T}$ is
the Thomson cross section, $m_{e}$ is the electron mass, and $X = 0.76$
is the primordial hydrogen abundance. Further, the gas-pressure profile
$P_{\rm gas}(x)$ can be written by a gas-density profile $\rho_{\rm
gas}(x)$ and a gas-temperature profile $T_{\rm gas}(x)$, as, 
\begin{equation}
P_{\rm gas}(x) = \frac{3 + 5X}{4} \frac{\rho_{\rm gas}(x)}{m_{p}}
                 k_{\rm B} T_{\rm gas}(x),
\end{equation}
where $m_{p}$ is the proton mass and $k_{\rm B}$ is the Boltzmann constant.
Thus we need the gas-density profile $\rho_{\rm gas}(x)$ and the
gas-temperature profile $T_{\rm gas}(x)$ to calculate the angular power
spectrum of SZ effect. 

\subsection{Gas-pressure profile of halos}

In \citet{komatsu01}, the gas-density profile and gas-temperature
profile were derived on the basis of the following assumptions:
\begin{itemize}
\item hydrostatic equilibrium between gas pressure and gravitational
potential due to dark matter
\item gas density tracing the dark-matter density in the outer parts
of halos
\item constant polytropic equation of state for gas: 
$P_{\rm gas} \propto \rho_{\rm gas}^{\gamma}$
\end{itemize}
The density profile of the dark matter is assumed to be the universal 
profile \citep{navarro97}:
\begin{equation}
\rho_{\rm DM}(x) = \frac{\rho_{\rm s}}{x (1+x)^{2}},
\end{equation}
where $\rho_{\rm s}$ is a scale density. Although, as \citet{cen01}
discussed, the finite lifetime of dark matter would change the density
profile of galaxy-scale halo, it is expected that this is not the case
with cluster scale because clusters of galaxies tend to form quite
recently \citep{lacey93} and thus there is little time to decay dark
matter particles after a cluster is formed. 

Then the gas-density profile is given analytically as,
\begin{equation}
\rho_{\rm gas}(x) = \rho_{\rm gas}(0) \left[
1 - B \left\{ 1 - \frac{\ln{(1+x)}}{x} \right\}
\right]^{1/(\gamma - 1)},
\end{equation}
where $B$, which is a constant, and the polytropic index $\gamma$ are  
given in terms of the concentration parameter $c$. Finally the
normalization of the gas density is obtained by requiring that the gas
density is the dark-matter density times  $\Omega_{\rm b}(z) /
\Omega_{\rm m}(z,\Gamma)$ at the virial radius. Note that this ratio is
not time-independent because of the decay of dark matter. Thus the
gas-pressure profile is completely determined. Here we must mention that
hydrostatic equilibrium cannot be realized in a strict sense in the
presence of finite decay width of dark matter. However, since the
lifetime we are interested in here is about the age of universe, this
approximation would remain valid. 

\section{SZ POWER SPECTRUM WITH DECAYING DARK MATTER}

We can calculate the angular power spectrum of SZ effect by the method
described in the previous section and the Press-Schechter mass function.
Nonzero decay width of the dark matter affects it in various ways.
The following two statement will help understand its feature. First,
since the dark matter decreases with time, the evolution of the comoving
number density of massive cluster is milder than that with stable dark
matter \citep{oguri03}. Second, due to the decrease of matter component
and the increase of the radiation component, the expansion law of the
universe changes. This decreases the comoving volume and the
angular-diameter distance at a given redshift compared with that of
universe with stable dark matter.  

\subsection{Angular power spectrum\label{sec:power}}

\begin{figure}
\includegraphics[width=84mm]{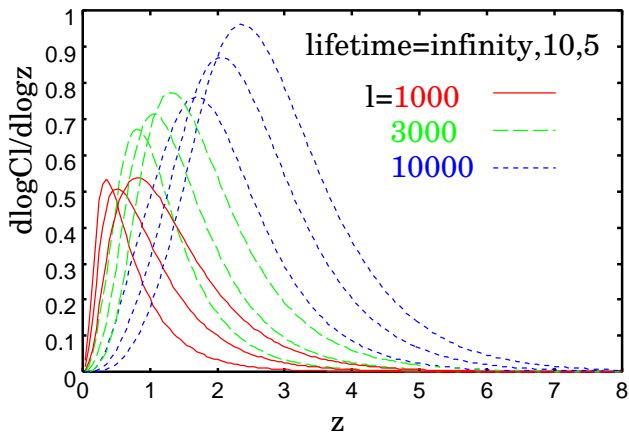}
\caption{Redshift distribution of $C_{l}$ for a given $l$: 
$d\ln{C_{l}}/dz$. Solid, dashed and dotted lines correspond to
$l=1000,3000,10000$, respectively, and for each $l$, three lines,
from the left to right, show model with the dark matter with lifetime 
of infinity, $10 h^{-1} {\rm Gyr}$ and $5 h^{-1} {\rm Gyr}$.
\label{fig:dCldz}} 
\end{figure}

It is helpful to see the contribution to $C_{l}$ of cluster with specific 
mass and redshift. Figure \ref{fig:dCldz} shows the redshift distribution
of $C_{l}$ for a given $l$: $d\ln{C_{l}}/dz$. Solid, dashed and dotted
lines correspond to $l=1000,3000,10000$, respectively, and for each $l$,
three lines, from the left to right, show model with the dark matter 
with lifetime of infinity, $10 h^{-1} {\rm Gyr}$ and $5 h^{-1} {\rm Gyr}$. 
It can be easily understood that clusters at a higher redshift contribute
to power of a larger $l$, that is, a smaller scale. Since the
angular-diameter distance at a given redshift become shorter with
shorter lifetime of the dark matter, the peaks move toward high-redshift
side as the lifetime becomes small for each $l$. 

\begin{figure}
\includegraphics[width=84mm]{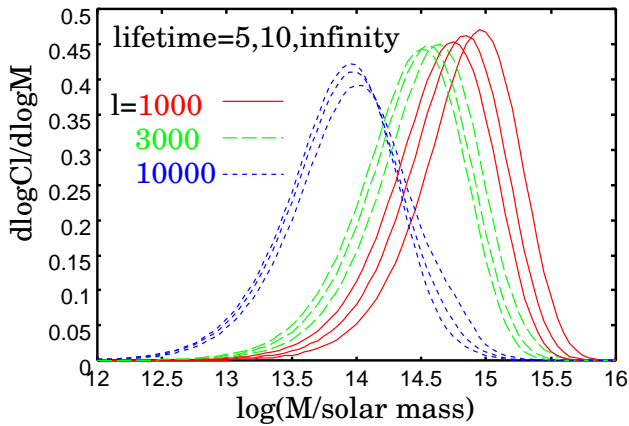}
\caption{Mass distribution of $C_{l}$ for a given $l$: 
$d\ln{C_{l}}/d\ln{M}$. Solid, dashed and dotted lines correspond to
$l=1000,3000,10000$, respectively, and for each $l$, three lines,
from the left to right, show model with the dark matter with lifetime 
of , $5 h^{-1} {\rm Gyr}$, $10 h^{-1} {\rm Gyr}$ and infinity.
Note that the order of the lifetime is opposite to Figure \ref{fig:dCldz}.
\label{fig:dCldM}} 
\end{figure}

Figure \ref{fig:dCldM} shows the redshift distribution of $C_{l}$
for a given $l$: $d\ln{C_{l}}/d\ln{M}$. Types of lines for each $l$ are
the same as those in Figure \ref{fig:dCldz} but the order of the
lifetime is opposite to that in Figure \ref{fig:dCldz}. It can be easily 
understood that clusters of larger mass contribute to power of a smaller
$l$, that is, larger scale. The peaks move toward low-mass side with
shorter lifetime because the angular-diameter distance at a given
redshift become shorter.

\begin{figure}
\includegraphics[width=84mm]{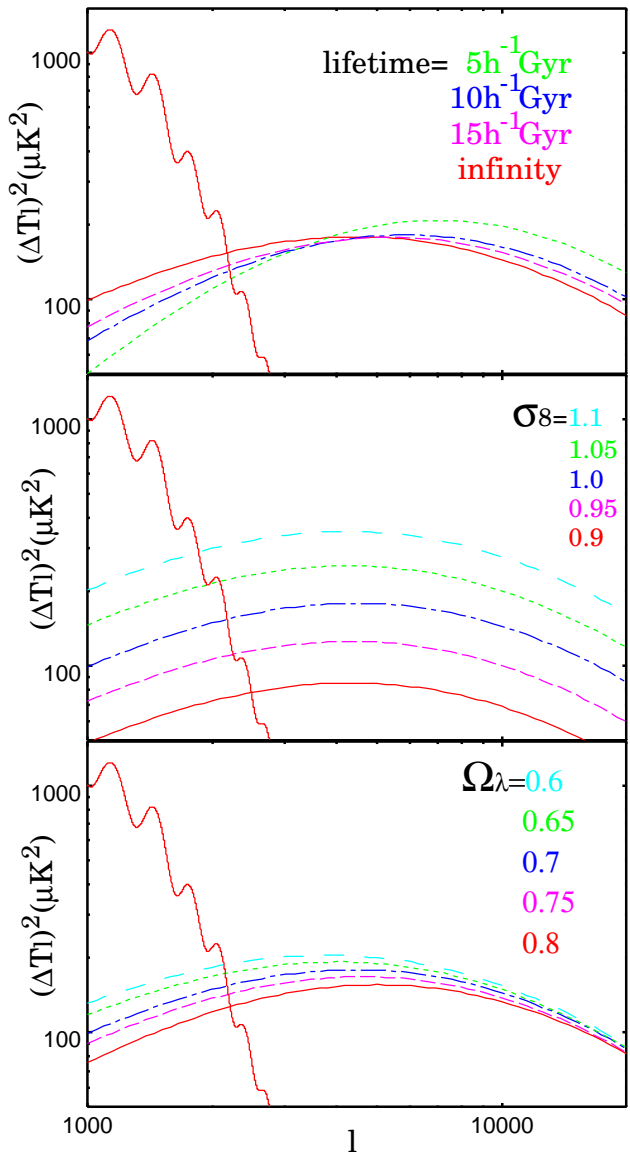}
\caption{Dependence of the angular power spectrum
on the lifetime of the dark matter, $\sigma_{8}$ and $\Omega_{\lambda}$.
The fiducial model is the universe with stable dark matter,
$\sigma_{8} = 1.0, \Omega_{\lambda} = 0.7$ and $h = 0.7$. The primary 
spectrum is also shown. 
\label{fig:Cl_three}} 
\end{figure}

In Figure \ref{fig:Cl_three}, we show the dependence of the angular
power spectrum on several cosmological parameters. Besides the lifetime
of dark matter, we choose $\sigma_{8}$ and $\Omega_{\lambda}$ as
main parameters. The reasons are (1) it is known that the SZ angular
power spectrum is quite sensitive to $\sigma_8$, and rather
insensitive to other parameters \citep{komatsu02}, (2) the SZ angular
power spectrum is also sensitive to baryon matter density $\Omega_{\rm
b}h^2$, where $h$ is the Hubble constant in units of $100{\rm
km\,s^{-1}Mpc^{-1}}$, but now it is accurately measured by CMB
anisotropy \citep{spergel03}, and (3) $\Omega_{\lambda}$ usually
degenerates with the lifetime of dark matter \citep{ichiki03,oguri03},
thus we should see how we can determine the lifetime of dark matter and
$\Omega_{\lambda}$ separately. The fiducial model here is the universe
with stable dark matter, $\sigma_{8} = 1.0$, $\Omega_{\lambda} = 0.7$,
$\Omega_{\rm b}h^2=0.02$, and $h = 0.7$. As can be seen in the top figure,
the power at large $l$ increase and that at small $l$ decrease with
shorter lifetime. This behavior is mainly because of the milder
evolution of the cluster abundance in decaying cold dark matter model.
The signal for a given multipole $l$ is from clusters which angular
sizes roughly corresponds to $l^{-1}$, so nearby clusters contribute
low-$l$ (see Figure \ref{fig:dCldz}). Since the decay of dark matter
lowers the ratio of low-$z$ to high-$z$ clusters, it tilts the angular
power spectrum so that signals at large $l$ increase. On the other hand,
the value of $\sigma_{8}$ increases or decreases the normalization of
the angular power spectrum and does not change the shape. The effect of
changing $\Omega_{\lambda}$ is also nearly scale-independent. From these
different dependences, we expect that we can determine these three
parameters rather separately.

\begin{figure}
\includegraphics[width=84mm]{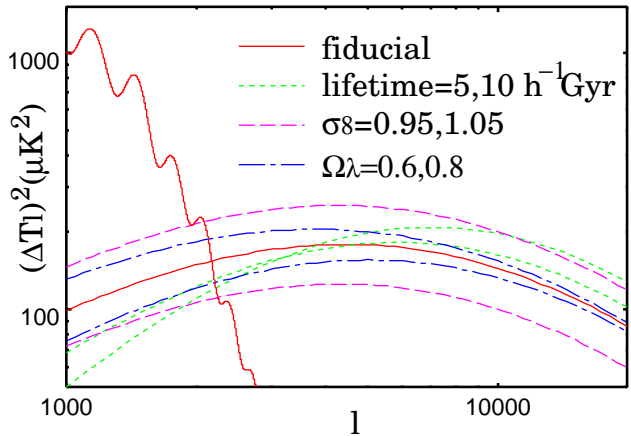}
\caption{Comparison of the strength of the dependence of
the lifetime, $\sigma_{8}$ and $\Omega_{\lambda}$ on the power spectrum.
The fiducial model is the same as in Figure \ref{fig:Cl_three}.
The primary spectrum is also shown.
\label{fig:Cl}} 
\end{figure}
 
The comparison of the strength of the dependence of each parameter is
shown in Figure \ref{fig:Cl}. From this Figure, we find that the
difference of signals by changing the lifetime by a factor 2 roughly
corresponds to the change of $\sigma_8$ by a few percents  or
$\Omega_\lambda$  by a few tens percents.

\subsection{Sensitivity of future experiment}

In this subsection we investigate how well we can probe the lifetime
of dark matter with future SZ surveys. We do not specify the
instruments but consider a survey with $100 {\rm deg}^{2}$ field of view,
beam size of $1'$ and sensitivity $2 \mu {\rm K}$. These are typical
values for near-future experiments such as ACT, AMIBA and BOLOCAM
\citep{komatsu02,AMIBA,BOLOCAM}.

\begin{figure}
\includegraphics[width=84mm]{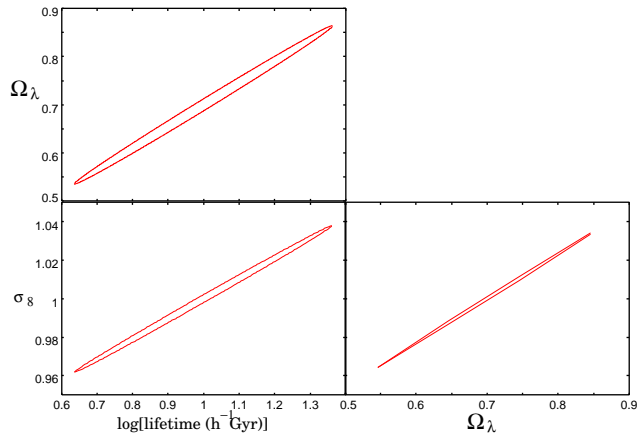}
\caption{Marginalized 1-$\sigma$ contour for $\Gamma, \sigma_{8}$, 
and $\Omega_{\lambda}$ for a cosmological model with 
$\Gamma^{-1} = 10 h^{-1} {\rm Gyr}$, $\sigma_{8} = 1.0$, and  
$\Omega_{\lambda} = 0.7$.
\label{fig:contour}} 
\end{figure}

To investigate the sensitivity for cosmological parameters, we use the
Fisher information matrix approach
\citep{seljak96,jungman96,zaldarriaga97,eisenstein99}. Because this
approach is based on the dependence of power spectrum on the
cosmological parameters, results would be rather insensitive to the
underestimation of halo abundance due to the Press-Schechter mass function.
Under the assumption of Gaussian perturbations and Gaussian noise, the
Fisher matrix for CMB anisotropies is, 
\begin{equation}
F_{ij} = \sum_{l} \frac{\partial C_{l}}{\partial p_{i}}
({\rm Cov}_{l})^{-1} \frac{\partial C_{l}}{\partial p_{j}},
\end{equation}
where $p_{i}$ is a cosmological parameter and the covariance matrix
${\rm Cov}_{l}$ is,
\begin{equation}
{\rm Cov}_{l} = \frac{2}{(2 l + 1) f_{\rm sky}}
\left[ C_{l} + 
\theta_{\rm beam}^{2} \sigma^{2} 
\exp{\left\{ \frac{l(l+1)}{8 \ln{2}} \theta_{\rm beam}^{2} \right\}}
\right].
\end{equation}
Here $f_{\rm sky}$ is a fraction of the surveyed sky, $\theta_{\rm beam}$
is the full width, half-maximum of the beam in radians, and $\sigma$ is
sensitivity in $\mu {\rm K}$. We use the values stated at the beginning
of this subsection for these parameters. Summation with respect to $l$
is taken from $l = 3000$ to $l = 20000$, in order for the primary power
to be negligible. As discussed in \S \ref{sec:power}, we adopt the
lifetime $\Gamma^{-1}$, $\sigma_{8}$, and $\Omega_{\lambda}$ as main
parameters which we vary. We fix the Hubble constant to $h=0.7$ and the
baryon density $\Omega_{\rm b}h^2=0.02$. Figure \ref{fig:contour} is
marginalized 1-$\sigma$ contour for these three parameters for a
cosmological model with $\Gamma^{-1} = 10 h^{-1} {\rm Gyr}$, $\sigma_{8}
= 1.0$, and $\Omega_{\lambda} = 0.7$. As can be seen, the lifetime can
be determined by a factor of two or so. On the other hand, $\sigma_{8}$
and $\Omega_{\lambda}$ will be determined within $\pm 0.04$ and $\pm
0.15$, respectively. 

It should be noted that there are some theoretical uncertainties 
in the SZ power spectrum. \citet{komatsu02} found that uncertainties due
to the mass function, concentration parameter, outer radius of the gas
profile and the effect of temperature decrease within $5\%$ of the virial
radius are negligible. \citet{dasilva01b} showed by hydrodynamical
simulations that the effect of non-adiabatic physics, such as radiative
cooling and preheating, reduces $C_{l}$ by 20-40$\%$ on all angular scales. 
This seems a significant uncertainty because the effect of finite
lifetime is not so large. However, since this uncertainty does not change
the shape of the spectrum but just changes the normalization, it is
expected that it does not weaken the constraint on the lifetime while
it affects the determination of $\sigma_{8}$ and $\Omega_{\lambda}$.
Of course, we need more accurate mass functions when we compare the
theoretical predictions with real observational data and constrain the
lifetime. 

\section{SUMMARY}

In this paper, we studied the effect of finite lifetime of dark matter
on SZ power spectrum. We calculated the SZ power spectrum following the
method proposed by \citet{komatsu02} using the Press-Schechter mass
function and taking the decay width of dark matter into account. We have
found that the finite lifetime of dark matter decreases the power at
large scale ($l < 4000$) and increases at small scale ($l > 4000$).
This unique feature will allow us to probe the lifetime of dark matter,
rather independently with $\sigma_8$ and $\Omega_{\rm m}$ which mainly
change the normalization of the angular power spectrum. Then we
estimated how well we can constrain the lifetime in the future CMB
experiment such as  ACT, AMIBA and BOLOCAM. An SZ survey with $100 {\rm
deg}^{2}$ field of view, beam size of $1'$ and sensitivity $2 \mu {\rm
K}$, which are the representative values of the near-future experiment,
will be able to determine the lifetime within factor of two if the
lifetime is $10 h^{-1} {\rm Gyr}$ even if we marginalized other
parameters such as $\sigma_8$ and $\Omega_{\rm m}$. Therefore, future SZ
surveys will definitely provide an opportunity to reveal the nature of
dark matter.

\section*{Acknowledgments}

We thank Takeshi Kuwabara for useful comments on observation of SZ effect,
Kei Kotake and Hiroshi Ohno for useful discussion. K.T.'s work is
supported by Grant-in-Aid for JSPS Fellows, and K.I.'s work is supported
in part by the Sasakawa Scientific Research Grant from the JSS.

\bsp

\label{lastpage}

\end{document}